\documentstyle[prl,aps,graphicx]{revtex}
\newcommand{\be}{\begin{equation}} \newcommand{\ee}{\end{equation}}
\newcommand{\ba}{\begin{array}} \newcommand{\ea}{\end{array}}
\newcommand{\ath}{\mathop{\rm arctanh}\nolimits}
\newcommand{\ach}{\mathop{\rm arccosh}\nolimits}
\begin{document} \draft 
\title{Gravitational potential and energy of homogeneous rectangular
parallelepiped}
\author{Zakir F. Seidov, P.I. Skvirsky}
\address{\it Department of Physics, Ben-Gurion University of the Negev,
Beer-Sheva, 84105, Israel}\maketitle \begin{abstract}
Gravitational potential and gravitational energy are presented in
{analytical} form for homogeneous right parallelepiped. \end{abstract}
\pacs{PACS numbers: 01.55.+b, 45.20.Dd,  96.35.Fs}
{\em{Keywords:}} Newtonian mechanics; mass, size, gravitational fields
\section{Introduction: Basic formulae}
As it is well known, Newtonian gravitational potential, of homogeneous
body with constant density $\rho$,  
at point (X,Y,Z) is defined as triple integral over the body's volume:
\be \label{UXYZ} U(X,Y,Z)=G\rho \int \int \int u(X,Y,Z,x,y,z)\,dxdydz \ee
with \be u(X,Y,Z,x,y,z)={[(x-X)^2+(y-Y)^2+(z-Z)^2]}^{-1/2};\ee
here $G$ stands for Newtonian constant of gravitation.
In spite of almost 400-year-long attempts since Isaac Newton' times,
the integral (\ref{UXYZ}) is known in closed form (not in the series!) in
quite a few cases \cite{Dub}:\\
a) a piece of straight line, b) a sphere, c) an ellipsoid;\\
note that both cases a) and  b) may be considered as particular cases of c).
As to serial solution, the integral (1) is expressed in terms of the 
various kinds of series for external points outside the
minimal sphere containing the whole body (non-necessary homogeneous),
as well as for inner points close to the origin of
co-ordinates. However in this note we do not touch the problem
of serial solution and are only interested in {\it exact analytical}
solution of (\ref{UXYZ}).\\
Recently Kondrat'ev and Antonov \cite{KA} have obtained the  rather sofisticate analytical
formulas for the gravitational potential (and the gravitational energy) of some 
axial-symmetric figures,  namely homogeneous  lenses with spherical surfaces of different
radii. In forthcoming paper \cite{GREN} we present some new solutions for homogeneous
bodies of revolution. Here we present  the solution of Eq. (\ref{UXYZ}) for homogeneous
right (=rectangular) parallelepiped, (RP). The great simplification available in this case
is that boundaries of integration over each of variables $x,y,z$ are {\it fixed and 
thus independent on other two variables} \cite{Budak}. This allows to get 
results in the closed form in terms of elementary functions. To this end
we widely used Mathematica \cite{Wolf}.\\
Let RP with center at the origin of Cartesian co-ordinates have lengths of sides
$2a,2b,2c,$ so that inside the RP, $-a\leq x\leq a,-b\leq y\leq b,-c\leq z\leq c$ . Then
Eq. (1) may be rewritten in the following symmetrical form:
\be \label{symm} U(X,Y,Z)=G\rho
\int_{xm}^{xp}dx\int_{ym}^{yp}dy\int_{zm}^{zp}u(x,y,z)dz,\ee
\be u(x,y,z)={(x^{2}+y^{2}+z^2)}^{-1/2};\ee
\be \label{bounds} xm=-a-X,xp=a-X,\, ym=-b-Y,yp=a-Y,\, zm=-c-Z,zp=c-Z. \ee
Note that now the boundaries (\ref{bounds}) of integration in Eq.(\ref{symm})  depend on
$X,Y,Z,$ but this is not so "dangerous" as the case of dependence of boundaries
on other variables of integration.\\
Now we are ready to start calculation of integral (\ref{symm}) in analytical form.
Certainly final result should not depend on the order of integration [3] and
this may be used to check the calculation.\\ 
First, an integration over $z$ gives: \be \label{line} 
uz(X,Y,Z,x,y)=\int_{zm}^{zp} u(x,y,z) dz=[\ln (z+1/u)]_{z=zm}^{z=zp}. \ee
Here $u$ is the same function as in Eq. (4).\\
Second, an integration over $y$ gives: \be \label{rect}
uyz(X,Y,Z,x)= \left [ \left[ -x\arctan{y z u \over x}+y\ln (z+ 1/u)+
z \ln(y+1/u)\right]_{z=zm}^{z=zp}\right ]_{y=ym}^{y=yp}.\ee
And  integration over $x$ gives the final expression for the
gravitational  potential of homogeneous RP:
\be \label{fin} U(X,Y,Z)= G\rho \biggl[\biggl[\biggl[ v(x,y,z)+v(y,z,x)+v(z,x,y)
\biggr] _{z=zm}^{z=zp}\biggr]_{y=ym}^{y=yp}\biggr]_{x=xm}^{x=xp};\ee
\be v(x',y',z')=y'z' \ln [x'+{({x'}^2+{y'}^2+{z'}^2)}^{1/2}]-{{x'}^2\over 2}\arctan
{y'z'\over x'{({x'}^2+{y'}^2+{z'}^2)}^{1/2}}.\ee
Here  co-ordinates $X,Y,Z$ are as given in (5).
In Appendix we present the final result (8,9,5) in explicit form. \\
These formulas (\ref{fin},9,5) may look cumbersome and unpractical while in
fact they allow to easily (and exactly!) calculate gravitational potential of homogeneous
RP at {\it arbitrary} point {\it inside} as well {\it outside} the body. The payment for
this universality does not seem very high. Also we should note that  obtaining the
result (\ref{fin},9,5) is non-trivial in the sense that direct using of Mathematica's
command for integration of (1,2), Integrate[f[x],x], (and especially command for
definite integral, Integrate[f[x],\{x,xm,xp\}]), gives enormous complex expressions even
leaving some integrals uncalculated!\\ 
Note that a physically evident symmetry relative to   all three sign changes,
$X\rightarrow -X$, $Y\rightarrow -Y$, and $Z\rightarrow-Z$ may be used to check final
formulas.\\Let us consider now the various particular cases.
\section{A piece of straight line} This is one of a few known classical results.
From Eq.(\ref{line}), the 3D-potential, at arbitrary point $(X,\,Y,\,Z)$, of a piece of
straight line, PSL, with constant linear density $\rho$ and length 2c, with center at
origin of co-ordinates , PSL laying along the $Z$-axis, is as follows:
  \be\label{poline} U_{line}(X,Y,Z)=G\rho\ln\biggl[{c-Z+{(X^2+Y^2+(c-Z)^2)}^{1/2}\over
-c-Z+{(X^2+Y^2+(c+Z)^2)}^{1/2}}\biggr].\ee
The PSL case may be considered as RP with two infinitesimal dimensions $dx,\,dy$.
 There is a scaling invariance: if we express all co-ordinates in units of $c$ then we
get the {\it universal} law coinciding with the potential of PSL with length equal to 2.
\section{Rectangle}
From Eq.(\ref{rect}) we may get  the 3D-potential, at arbitrary point $(X,\,Y,\,Z)$, of a
rectangular with constant surface density $\rho$, with sides of lengths 2c and 2b along
the $Z$ and $X$ axes respectively, with center at origin of co-ordinates $X,\,Y,\,Z$.
Equipotential surfaces are symmetrical relative to the rectangle's plane. 
The "rectangle" may be considered as RP with one infinitesimal dimension $dx$.
\section{Cube} From all RPs with three dimensions of RP being {\it finite}, 
the case of cube is of the utmost interest. We consider this case in detail.
First, the gravitational potential of homogeneous cube is symmetric relative to all three
co-ordinates which may have only even powers. Second, there is a scaling invariance:
if we express all co-ordinates in units of cube edge half-length, $a$, then we get the 
{\it universal} law coinciding with the gravitational potential of homogeneous cube
with edge length equal to 2. \subsection{Some particular points}
Here are values of gravitational potential of homogeneous cube, with density $\rho$ and
with edge length $2a$, at some particular points (note that here all potentials are given
in units of $a^2\,G\,\rho$):\\ at the center:
\be \label{p000} U(0,0,0)= 24 \ln{1+\sqrt{3}\over\sqrt{2}}-2\pi=9.52017;\ee
at the vertex: \be \label{p111} U(1,1,1)=12 \ln{\sqrt{3}+1\over\sqrt{2}}-\pi={1\over
2}U(0,0,0);\ee at the center of the face:
\be \label{p001} U(0,0,1)=4 \ath{157\over129}\sqrt{2\over 3}-
4 \arctan{7\over 3}\sqrt{2\over 3};\ee at the middle of the edge:
\be \label{p011} U(0,1,1)=4 \ln {10} -\arctan {4\over 3} -8 \arctan {1\over 3}=
4\ln {10}- \arctan {44\over117}-\pi.\ee
Though the  solution we discuss is exact however it is rather complex and
is difficult to use, so in the next sections we present some serial expansions.
\subsection{Neighborhood of the center}\label{nc}
 For the homogeneous cube with density $\rho$ and length of edge $2a$, the gravitational 
potential near the center is sphericall-symmetric and quadratic in co-ordinates (note
that all co-ordinates are in this paragraph expressed in units of $a$). The
deviation from spherical symmetry appears in the terms of fourth and higher orders:
\be \label{cntr} \begin{array}{l} U(X,Y,Z)\equiv U_s=a^2 G\rho \biggl[
\{24 \ln {1+\sqrt{3}\over\sqrt{2}}-2 \pi\}-\{{2\over 3} \pi
(X^2+Y^2+Z^2)\}+\\ \\  \{-{4\over 9\sqrt{3}}(X^4+Y^4+Z^4)+{4\over 3
\sqrt{3}}(X^2 Y^2+X^2 Z^2+ Y^2 Z^2)\}+\{- 
{1\over 162 \sqrt{3}}(X^6+Y^6+Z^6)+\\  \\ {5\over 108 \sqrt{3}}(X^2 Y^4 +
X^2 Z^4+X^4 Y^2+ X^4 Z^2+Y^2 Z^4 + Y^4 Z^2)-
{5\over 9 \sqrt{3}}X^2 Y^2 Z^2\}\biggr].\end{array} \ee 
Here inner braces separate the terms of zeroth, second, fourth and sixth order
respectively. Note that no terms with odd powers may occur in serial expansion of the
potential of cube and more generally the potential of RP.
\subsection{Potential at main diagonal of cube} The serial expansion of the gravitational
potential at the main diagonals where
abs$(X)=$abs$(Y)=$abs$(Z)=r\,\sqrt{3}$, and $r=\sqrt{(X^2+Y^2+Z^2)}$ is distance from the
center) of the homogeneous cube up to $r^{20}$ is as follows:
\be \label{diag} \begin{array}{l} W_{diag}(r)=a^2\,G\,\rho\,
\sum_{i=0}^{i=10}c_{2i}\biggl({r^2\over 3}\biggr)^i;
c_0=24 \ln {1+\sqrt{3}\over\sqrt{2}}-2 \pi,  
c_2=-2 \pi,\,c_4={8\over 3\,\sqrt{3}},\\ \\
c_6=-{8\over 27\,\sqrt{3}}, c_8={136\over 567\,\sqrt{3}},
c_{10}=-{104\over 2187\,\sqrt{3}},c_{12}={54392\over 1082565\,\sqrt{3}},
c_{14}=-{136\over 19683\,\sqrt{3}},\\ \\
c_{16}={842168\over 55801305\,\sqrt{3}},\,c_{18}=-{8632\over 4782969\,\sqrt{3}},\,
c_{20}={11003576\over 1363146165\,\sqrt{3}}.\end{array} \ee
This series represent the actual potential at the main diagonal very accurately; at the
"final" point, at the vertex, at $r=\sqrt{3}$, the difference between series and exact
solution ($4.76016$) is only $-0.00612$. \subsection{Potential at co-ordinate axis }
Taking $Y=0,\,Z=0$ we get potential at $X$ axis which  connects the centers of opposite
sides. Even in this case exact expression is rather complex, so we present the serial
expansion around the center (at point $X=0$). Series up to $X^{20}$ is as follows:
\be \label{UX} \begin{array}{l} U(X,0,0)=a^2\,G\,\rho
\sum_{i=0}^{i=10}c_{2i}X^{2i}; c_0= 24\,\ln{(1 + \sqrt{3})\over \sqrt{2}}-2\,\pi,
c_2=-{2\,\pi\over 3},\,c_4=-{4\over 9\,\sqrt{3}},\\ \\
c_6=-{1\over 162\,\sqrt{3}},c_8={17\over 1701\,\sqrt{3}},
c_{10}={13\over 104976\,\sqrt{3}},c_{12}=-{4307\over 6495390\,\sqrt{3}},
c_{14}=-{17\over 5668704\,\sqrt{3}},\\ \\
c_{16}={174431\over 2678462640\,\sqrt{3}},\,c_{18}={229\over 7346640384\,\sqrt{3}},
c_{20}=-{346867\over 43620677280\,\sqrt{3}}.\end{array}\ee
\subsection{Approximations by sphere} There are many ways of comparing the
cube, with edge
length $2a$ and density $\rho$, with a  homogeneous sphere.\\
1.First naive approximation is by the "inscribed" sphere of radius $R_1=a$ and with  the
same {\em density} $\rho$.  This gives the next rough estimation for the gravitational
energy of cube (here $M_1=4\pi/3\rho R_1^3$ is the mass of sphere): 
$$W_1={3\over 5}{G\,M_1^2\over R_1}={16\over 15}{\pi}^2G\,{\rho}^2\,a^5.$$
Note that gravitational potential energy $W$ of any body is of negative sign but in this
paper we loosely write all $W$'s with positive sign.\\
2. Second approximation is by the "inscribed" sphere   with the {\it mass} equal to cube's
mass (density of this sphere will be larger than $\rho$, density of cube). Then 
we have the next, less rough, estimation for the gravitational   energy of cube (here
$M_2=8\rho a^3$ is the mass of sphere): 
$$W_2={3\over 5}{G\,M_2^2\over a}={192\over 5}G\,{\rho}^2\,a^5.$$
3.Next approximation is by the sphere whose volume equal to cube's volume. 
Radius $R_3$ of this "equivolume" sphere is $R_{3}=(6/\pi)^{1/3}a$. Homogeneous sphere
with this radius and with the  density $\rho$ equal to density of cube gives the next
approximation for the gravitational energy of cube (here $M_3=8\rho a^3$ is the mass of
sphere): \be \label{W3} W_3={3\over 5}{G\,M_3^2\over R_3}={192\over 5}({\pi
\over 6})^{1/3}G\,{\rho}^2\,a^5=30.95\,G\,{\rho}^2\,a^5.\ee
Gravitational potential at the center of such sphere, $2\pi\,G\,\rho
R_3^2=2\pi({6/\pi})^{2/3}G\rho a^2=9.672 \,G\rho a^2$  differs from the 
exact value of central potential of cube (see Eq. (\ref{p000}))  by less than  
$1.6\%$. We deduce from this that $W_3$ (\ref{W3}) should be rather good
approximation to gravitational energy of homogeneous cube.\\
4.Radius $R_{4}$ of sphere with the same density and the same  central potential as ones
of the cube is $[U(0,0,0)/2\pi]^{1/2}a$, with  $[U(0,0,0)$ given by (\ref{p000}). This
gives another good estimation of gravitational energy of cube:
\be \label{W4} W_4={3\over 5}{G\,M_4^2\over R_4}={16\over
15}{\pi}^{2}\biggl({U(0,0,0,)\over 
2\pi}\biggr)^{5/2}\,G\,{\rho}^2\,a^5=29.75\,G\,{\rho}^2\,a^5.\ee
To calculate the potential energy of homogeneous cube   we used Mathematica to integrate
numerically the formula (\ref{fin}) over the volume of the cube and obtained the numerical
value \be\label{Wcu} W_{cube}=30.117\,G\rho^2a^5.\ee  We note that  (\ref{W3}) and
(\ref{W4}) give very accurate bounds to "exact" value (\ref{Wcu}).
\section{Potential at the center of RP} We write down the potential in the center of RP:
\be \label{URP0} \ba{l}U_{RP}(0,0,0)=4*\{ a b \ln {d+c\over d-c}+ b c \ln {d+a\over d-a}+
 c d \ln {d+b\over d-b}-\\ a^2 \arctan {b c \over a d}- b^2 \arctan {a c \over b d}-
 c^2 \arctan {a b \over c d}\}.\ea \ee
Here $d=(a^2+b^2+c^2)^{1/2}$ is the main diagonal of RP. Even potential in
the center of RP could not be scaled by values of $a,b,c$.
Three particular cases are of the larger interest:\\
a){\it cube} corresponding to case {$c=b=a$}, see (\ref{p000});\\
b){\it long thin stick with square cross-section} corresponding to case $a>>b=c$:
\be \label{Ust} U_{stick}(0,0,0)=b^2(-8\,\ln(b/a)+12-2\,\pi+4\,\ln 2);\ee
c){\it thin square plate} corresponding to the case $a<<b=c$:
\be \label{Upl} \,U_{plate}(0,0,0)=16\,a\,b\,\ln (\sqrt{2}+1)-2\,\pi \,a^2.\ee
These formulae will be used further for comparing RP with ellipsoid.
\section{Comparison with ellipsoid}
It is interesting to compare the homogeneous RP and the homogeneous triaxial ellipsoid 
with semi-axes  $A,\,B,\,C\,$, and with  density and   central value of gravitational
potential as ones of RP. \\The potential in the center of homogeneous triaxial ellipsoid
with semi-axes $A,\,B,\,C$ and density $\rho$ is \cite{LaLi} :
\be \label{Uel} U_{ell}(0,0,0)=\pi\,\rho\,G\,A\,B\,C\,\int_0^\infty{d\,s\over
\sqrt{(A^2+s)(B^2+s)(C^2+s)}}.\ee The potential energy of triaxial ellipsoid  is
\cite{LaLi}: \be \label{Wel} W_{ell}={3\over 10}G\,M^2\int_0^\infty{d\,s\over
 \sqrt{(A^2+s)(B^2+s)(C^2+s)}}.\ee
Here $M=4/3\,\pi\,\rho\,A\,B\,C $ is the ellipsoid's mass.\\ From (\ref{Uel}) and
(\ref{Wel}), we note the interesting relation between the gravitational potential at the
center of homogeneous ellipsoid and the gravitational potential energy of the ellipsoid:
\be \label{WUMel} W_{ell}={2\over 5}U_{ell}(0,0,0)\,M_{ell} \ee for any semi-axes!\\
For the cube we have from  (\ref{p000}) and (\ref{Wcu}) the relation  \be \label{WUMcu}
W_{cube}=0.39544\,U_{cube}(0,0,0)\,M_{cube}, \ee which is very close to (\ref{WUMel}).\\
As integral in (\ref{Uel}) and (\ref{Wel}) in general case is expressed only in terms of
elliptic integrals, to simplify calculations we assume $A>B=C$ and also $a>b=c$, then we
have for the potential energy of ellipsoid of revolution \cite{LaLi}:
\be W_{ell}={3\over5}{GM^2\over\sqrt(A^2-B^2)}\ach{A\over B}.\ee
and for potential at the center:
\be U_{ell}(0,0,0)={5\over 2\,M_{ell}} W_{ell}.\ee
Here $M=4/3\,\pi\,\rho\,A\,B^2 $ is the mass of the ellipsoid of revolution.\\
There is no simple relation between $U(0,0,0)$ and $W$ for RP with arbitrary 
values of edge lengths and we may use relation (\ref{WUMel}) for rather precise 
estimation of potential energy of RP with known $U(0,0,0)$ from (\ref{URP0}).
\subsection*{Potential at vertex of RP}  Potential of RP at vertex  is:
$ U(A,B,C)=1/2\,U(0,0,0)$ that is exactly half of the potential at center of RP! 
\section{Potential energy of RP}
To get the potential energy of RP, $WRP$, we need to calculate triple
integral over the volume of body:
\be WRP={1\over 2} \rho \int_{-a}^a dX \int_{-b}^b dY \int_{-c}^c
U(X,Y,Z,a,b,c)dZ. \ee After some lengthy interactive session with Mathematica, we
get  the potential energy of homogeneous RP with density $\rho$ and edge lengths $2
a,\,2 b,\, 2 c$, which we write down  in the following concise form:
\be \label{WRP} \ba {l}\Huge WRP=G\,\rho^2 [f(a,b,c)+f(b,c,a)+f(c,a,b)];
f(a,b,c)=c_5 a^5+c_4 a^4+c_3 a^3+c_2 a^2; \\
 c_5={32\over 15};\,\,c_4={32\over 15}(d-d1-d3)-
{16\,b\over 3}\ln{(d1-b)(d+b)\over a\,d3}-
{16\,c\over 3}\ln{(d3-c)(d+c)\over a\,d1};\\
c_3=-{64\, b\, c\over 3}\arctan {b\, c\over a\,d};\,d=\sqrt{a^2+b^2+c2};\,
d1=\sqrt{a^2+b^2};\,d3=\sqrt{a^2+c^2};\\ \\
c_2={32\,b^2 \over 5} ( d1 -d)  + {32\,c^2  \over 5} ( d3 -d )  - 
 16\,b\,{c^2}\,\ln {d-b\over d+b }-16\,{b^2}\,c\,\ln {d-c\over d+c}.\ea\ee
\subsection{Potential energy of cube}
From (\ref{WRP}), taking $c=a,\,b=a$, we get the potential energy of homogeneous
cube with edge length $2 a$:
\be WC=32 G\rho^2 a^5 \{{2\sqrt{3}-\sqrt{2}-1\over 5}+{\pi \over3}+
\ln [(\sqrt{2}-1)(2-\sqrt{3})]\}=30.117 G \rho^2 a^5. \ee 
\subsection{Potential energy of thin long stick}
We consider a case $a>>b=c$ that is a case of thin long stick with square cross-section. 
Leading term in expansion of WRP (\ref{WRP}) gives the 
potential energy of thin long stick:
\be \label{Wst} W_{stick}={32\over 3}G \rho^2 a\,b^4\ln{a\over b}. \ee
From this and (\ref{Ust}) we get for stick:
\be {W_{stick}\over 8\,a\,b^2 U_{stick}(0,0,0)}={1\over 2} .\ee
\subsection{Potential energy of thin square plate}
Taking one of RP dimension infinitesimally small, $a\rightarrow 0$, we get, 
from Eq.(\ref{WRP}), the potential energy of
thin rectangular plate. Note that first non-zero term in expansion is
{\it quadratic} in $a$. If we additionally take $b=c$, then we get the
potential energy of thin square  plate:
\be \label{WS} WS=64\,G\,\rho^2b^3\,a^2\,\left(\ln (\sqrt{2}+1)-
{\sqrt{2} -1 \over 3}  \right) =47.5714\, G\,\rho^2b^3 a^2.\ee
From (\ref{Upl}) and (\ref{WS}) we have another limit for relation WUM:
\be {WS\over 8\,a\,b^2 U_{pl}(0,0,0)}=
{1\over 2}-{\sqrt{2}-1\over 6\,\ln(\sqrt{2}+1)}=.421673.\ee
\subsection{Relation between potential energy, gravitational potential
and mass of RP}
General picture  of relation between potential energy, gravitational potential
and mass of RP is shown in the Fig.1. We shortly denote it as $WUM$ which means:\\
"potential energy $W$/(potential at the center $U(0)\,$ x mass of RP $M$)".\\
For homogeneous ellipsoid $WUM=2/5$, see (\ref{WUMel}).  
\begin{figure}  \includegraphics[scale=.6]{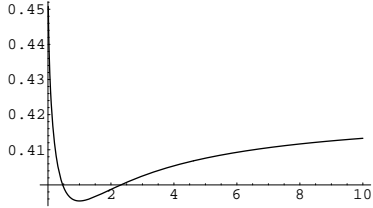}
\caption{WUM for RP with square cross section $2\,b$x$2\,b$ and length $2\,a$.
Abscissas are values of $b/a$ and ordinates are values of relation $WUM=W/U(0)\,M$ that is
ratio of gravitational potential energy of RP to product of gravitational  potential at
the center of RP by mass of homogeneous RP. $WUM$ has minimum for cube ($b/a=1$), tends to
1/2 at $b/a\rightarrow 0$ (thin long stick) and to
${1\over 2}-{\sqrt{2}-1\over 6\,\ln(\sqrt{2}+1)}=.421673$ at 
$b/a\rightarrow \infty$ (thin square plate).
For homogeneous ellipsoid $WUM=2/5$.} \end{figure} \section*{Acknowledgements} We are
grateful to  E. Liverts for valuable discussions and to M. Trott for useful
correspondence. This work was partly funded by Israel Ministry of Absorbtion. 	
 
\section*{Appendix} Here we present a full explicit expression for gravitational potential
of homogeneous rectangular parallelepiped with density $\rho$ and with lengths of  edges
$a,\,b,\,c$ along $X,\,Y\,$ and $Z\,$ axes respectively; $X,\,Y,\,Z$  are Cartesian
co-ordinates of point at which the gravitational potential is calculated, and the origin
of co-ordinates coincides with the center of parallelepiped. This expression is valid for
any point $X,\,Y,\,Z$ inside as well as outside the body: \begin{verbatim}
U(X,Y,Z,a,b,c)=
{(-c-Z)^2*ArcTan[((-a-X)*(-b-Y))/(Sqrt[(-a-X)^2+(-b-Y)^2+(-c-Z)^2]*(-c-Z))]- 
(-c-Z)^2*ArcTan[((a-X)*(-b-Y))/(Sqrt[(a-X)^2+(-b-Y)^2+(-c-Z)^2]*(-c-Z))]- 
(-c- Z)^2*ArcTan[((-a-X)*(b-Y))/(Sqrt[(-a-X)^2+(b-Y)^2+(-c-Z)^2]*(-c-Z))]+ 
(-c-Z)^2*ArcTan[((a-X)*(b-Y))/(Sqrt[(a-X)^2+(b-Y)^2+(-c-Z)^2]*(-c-Z))]+ 
(-b-Y)^2*ArcTan[((-a-X)*(-c-Z))/((-b-Y)*Sqrt[(-a-X)^2+(-b-Y)^2+(-c-Z)^2])]+ 
(-a-X)^2*ArcTan[((-b-Y)*(-c-Z))/((-a-X)*Sqrt[(-a-X)^2+(-b-Y)^2+(-c-Z)^2])]- 
(-b-Y)^2*ArcTan[((a-X)*(-c-Z))/((-b-Y)*Sqrt[(a-X)^2+(-b-Y)^2+(-c-Z)^2])]- 
(a -X)^2*ArcTan[((-b-Y)*(-c-Z))/((a-X)*Sqrt[(a -X)^2+(-b-Y)^2+(-c-Z)^2]]- 
(b-Y)^2*ArcTan[((-a -X)*(-c-Z))/((b-Y)*Sqrt[(-a-X)^2+(b-Y)^2+(-c-Z)^2])]- 
(-a-X)^2*ArcTan[((b-Y)*(-c-Z))/((-a-X)*Sqrt[(-a-X)^2+(b-Y)^2+(-c-Z)^2])]+ 
(b-Y)^2*ArcTan[((a-X)*(-c-Z))/((b-Y)*Sqrt[(a-X)^2+(b-Y)^2+(-c-Z)^2])]+ 
(a-X)^2*ArcTan[((b-Y)*(-c-Z))/((a-X)*Sqrt[(a-X)^2+(b-Y)^2+(-c-Z)^2])]- 
(c-Z)^2*ArcTan[((-a-X)*(-b-Y))/(Sqrt[(-a-X)^2+(-b-Y)^2+(c-Z)^2]*(c-Z))]+ 
(c-Z)^2*ArcTan[((a-X)*(-b-Y))/(Sqrt[(a-X)^2+(-b-Y)^2+(c-Z)^2]*(c-Z))]+ 
(c-Z)^2*ArcTan[((-a-X)*(b-Y))/(Sqrt[(-a-X)^2+(b-Y)^2+(c-Z)^2]*(c-Z))]- 
(c-Z)^2*ArcTan[((a-X)*(b-Y))/(Sqrt[(a-X)^2+(b-Y)^2+(c-Z)^2]*(c-Z))]- 
(-b-Y)^2*ArcTan[((-a-X)*(c-Z))/((-b-Y)*Sqrt[(-a-X)^2+(-b-Y)^2+(c-Z)^2])]- 
(-a-X)^2*ArcTan[((-b-Y)*(c-Z))/((-a-X)*Sqrt[(-a-X)^2+(-b-Y)^2+(c-Z)^2])]+ 
(-b-Y)^2*ArcTan[((a-X)*(c-Z))/((-b-Y)*Sqrt[(a-X)^2+(-b-Y)^2+(c-Z)^2])]+ 
(a-X)^2*ArcTan[((-b-Y)*(c-Z))/((a-X)*Sqrt[(a-X)^2+(-b-Y)^2+(c-Z)^2])]+ 
(b-Y)^2*ArcTan[((-a-X)*(c-Z))/((b-Y)*Sqrt[(-a-X)^2+(b-Y)^2+(c-Z)^2])]+ 
(-a-X)^2*ArcTan[((b-Y)*(c-Z))/((-a-X)*Sqrt[(-a-X)^2+(b-Y)^2+(c-Z)^2])]- 
(b-Y)^2*ArcTan[((a-X)*(c-Z))/((b-Y)*Sqrt[(a-X)^2+(b-Y)^2+(c-Z)^2])]- 
(a-X)^2*ArcTan[((b-Y)*(c-Z))/((a-X)*Sqrt[(a-X)^2+(b-Y)^2+(c-Z)^2])]}/2- 
(-b-Y)*(-c-Z)*Log[-a-X+Sqrt[(-a-X)^2+(-b-Y)^2+(-c-Z)^2]]- 
(-a-X)*(-c-Z)*Log[-b-Y+Sqrt[(-a-X)^2+(-b-Y)^2+(-c-Z)^2]]+ 
(-b-Y)*(-c-Z)*Log[a-X+Sqrt[(a-X)^2+(-b-Y)^2+(-c-Z)^2]]+ 
(a-X)*(-c-Z)*Log[-b-Y+Sqrt[(a-X)^2+(-b-Y)^2+(-c-Z)^2]]+ 
(b-Y)*(-c-Z)*Log[-a-X+Sqrt[(-a-X)^2+(b-Y)^2+(-c-Z)^2]]+ 
(-a-X)*(-c-Z)*Log[b-Y+Sqrt[(-a-X)^2+(b-Y)^2+(-c-Z)^2]]- 
(b-Y)*(-c-Z)*Log[a-X+Sqrt[(a-X)^2+(b-Y)^2+(-c-Z)^2]]-
(a-X)*(-c-Z)*Log[b-Y+Sqrt[(a-X)^2+(b-Y)^2+(-c-Z)^2]]+ 
(-b-Y)*(c-Z)*Log[-a-X+Sqrt[(-a-X)^2+(-b-Y)^2+(c-Z)^2]]+ 
(-a-X)*(c-Z)*Log[-b-Y+Sqrt[(-a-X)^2+(-b-Y)^2+(c-Z)^2]]- 
(-b-Y)*(c-Z)*Log[a-X+Sqrt[(a-X)^2+(-b-Y)^2+(c-Z)^2]]- 
(a-X)*(c-Z)*Log[-b-Y+Sqrt[(a-X)^2+(-b-Y)^2+(c-Z)^2]]- 
(b-Y)*(c-Z)*Log[-a-X+Sqrt[(-a-X)^2+(b-Y)^2+(c-Z)^2]]- 
(-a-X)*(c-Z)*Log[b-Y+Sqrt[(-a-X)^2+(b-Y)^2+(c-Z)^2]]+ 
(b-Y)*(c-Z)*Log[a-X+Sqrt[(a-X)^2+(b-Y)^2+(c-Z)^2]]+ 
(a-X)*(c-Z)*Log[b-Y+Sqrt[(a-X)^2+(b-Y)^2+(c-Z)^2]]- 
(-a-X)*(-b-Y)*Log[-c+Sqrt[(-a-X)^2+(-b-Y)^2+(-c-Z)^2-Z]+ 
(a-X)*(-b-Y)*Log[-c+Sqrt[(a-X)^2+(-b-Y)^2+(-c-Z)^2]-Z]+ 
(-a-X)*(b-Y)*Log[-c+Sqrt[(-a-X)^2+(b-Y)^2+(-c-Z)^2]-Z]- 
(a-X)*(b-Y)*Log[-c+Sqrt[(a-X)^2+(b-Y)^2+(-c-Z)^2]-Z]+ 
(-a-X)*(-b-Y)*Log[c+Sqrt[(-a-X)^2+(-b-Y)^2+(c-Z)^2]-Z]- 
(a-X)*(-b-Y)*Log[c+Sqrt[(a-X)^2+(-b-Y)^2+(c-Z)^2]-Z]- 
(-a-X)*(b-Y)*Log[c+Sqrt[(-a-X)^2+(b-Y)^2+(c-Z)^2]-Z]+ 
(a-X)*(b-Y)*Log[c+Sqrt[(a-X)^2+(b-Y)^2+(c - Z)^2]-Z].\end{verbatim} 
 Here the potential is given in units of $G\,\rho\,$ and the Mathematica's language is
used except of first "additional" line,$U(X,Y,Z,a,b,c)=$;  Mathematica's designations are
corresponding to "usual" mathematical formulas as follows: Log[x]$\equiv \ln (x)$,
Arctan[x]$\equiv \arctan (x)$, Sqrt[x]$\equiv x^{1/2}$.\\
\end{document}